# Mean arc theorem for exploring domains with randomly distributed arbitrary closed trajectories

Samuel Hidalgo-Caballero[1,2], Alvaro Cassinelli[3], Matthieu Labousse[2] & Emmanuel Fort[1*]

[1]Institut Langevin, ESPCI Paris, Université PSL, CNRS, 75005 Paris, France, EU.

[2]Gulliver, UMR CNRS 7083, ESPCI Paris, Université PSL, 75005 Paris, France, EU.

[3] School of Creative Media City, City University of Hong Kong, 18 Tat Hong Ave, Kowloon Tong, Hong Kong.

Corresponding author: emmanuel.fort@espci.fr

**Abstract:**

A remarkable result from integral geometry is Cauchy's formula, which relates the mean path length of ballistic trajectories randomly crossing a convex 2D domain, to the ratio between the region area and its perimeter. This theorem has been generalized for non-convex domains and extended to the case of Brownian motion to find many applications in various fields including biological locomotion and wave physics. Here, we generalize the theorem to arbitrary closed trajectories exploring arbitrary domains. We demonstrate that, regardless of the complexity of the trajectory, the mean arc length still satisfies Cauchy's formula provided that no trajectory is entirely contained in the domain. Below this threshold, the mean arc length decreases with the size of the trajectory. In this case, an approximate analytical formula can still be given for convex trajectories intersecting convex domains provided they are small in comparison. To validate our analysis, we performed numerical simulations of different types of trajectories exploring arbitrary 2D domains. Our results could be applied to retrieve geometric information of bounded domains from the mean first entrance-exit length.

**Significance statement:** Integral geometry provides probabilistic strategies to measure geometrical quantities with multiple applications in stereology, shape-recognition, and nuclear reactors. One of its most remarkable results is the Cauchy theorem which states that the mean length of randomly distributed chords intersecting any 2D domain is proportional to its surface over its perimeter. We generalize this theorem to any arbitrary closed-looped trajectory, provided that no trajectory is entirely inscribed in the domain. Interestingly, an approximate formula can still hold beyond this limit. Closed trajectories are ubiquitous in many systems such as electrons in magnetic fields, active colloids, and robots. Our analysis also gives a criterion for predicting when an arbitrary shape fits completely into another arbitrary one, a crucial problem in packaging optimization.

# 1. Introduction

More than one and a half century ago Cauchy demonstrated a remarkable formula relating the mean length of chords randomly distributed intercepting an arbitrary convex domain with its geometry, namely its size and shape (1). The mean chord length was found to be proportional to the ratio between the area and perimeter in 2D, and the volume and area for 3D objects (2). This theorem has since been extended to convex hypervolumes in arbitrary Euclidean dimensions (3) and to non-Euclidian spaces too (4). This invariant property finds multiple applications in fields such as ecology (5), image analysis (6), stereology (7), reactor design (8, 9) and characterization of random media (10).

A significant step was taken when the theorem was also extended to trajectories associated to Brownian motion (11–14) with experimental implementations using living organisms such as insects and bacteria (15, 16). In this case, the geometric parameters of the domain are related to the mean trajectory length inside the domain, i.e., the first entrance-exit mean length. The theorem has even been generalized to the context of waves propagating in scattering media, both theoretically and experimentally (17, 18).

In this paper, we address the possibility of further generalizing the mean chord theorem to intersecting arcs associated to closed-looped trajectories, an interesting problem with practical applications since dynamics generating this type of trajectories are pervasive in natural or artificial systems, such as electrons moving in a magnetic field (19), swimming organisms (20–22), robots (23) and topologically protected states of transport (24). We show that in the case of arbitrary closed-loop trajectories, Cauchy's relation can be recovered from the mean arc length of homogeneously distributed random paths intersecting arbitrary 2D bounded domains. The validity of this theorem imposes that no trajectory is entirely enclosed in the explored domain. In the case of small trajectories, an approximate relation still holds relating the mean arc length to the shape and size of the domain. A corollary of this theorem can be used to prove the mutual embeddability between any arbitrary objects, i.e., the property of any arbitrary shape to be inscribed into one.

The article is organized as follows. First, we perform numerical simulations of the particular case of circular trajectories to better understand the evolution of the mean arc length as a function of loop size relative to the size of the explored domain. In a second part, we give the proof and the validity limit of the mean arc length theorem. Then, we perform numerical implementations with arbitrary trajectories and domain shapes to validate the mean arc length formula. Finally, we extend the theorem with an approximate formula in the regime of small trajectories relative to the explored domain and show the good agreement with numerical simulations.

## 2. The special case of circular trajectories exploring arbitrary domains

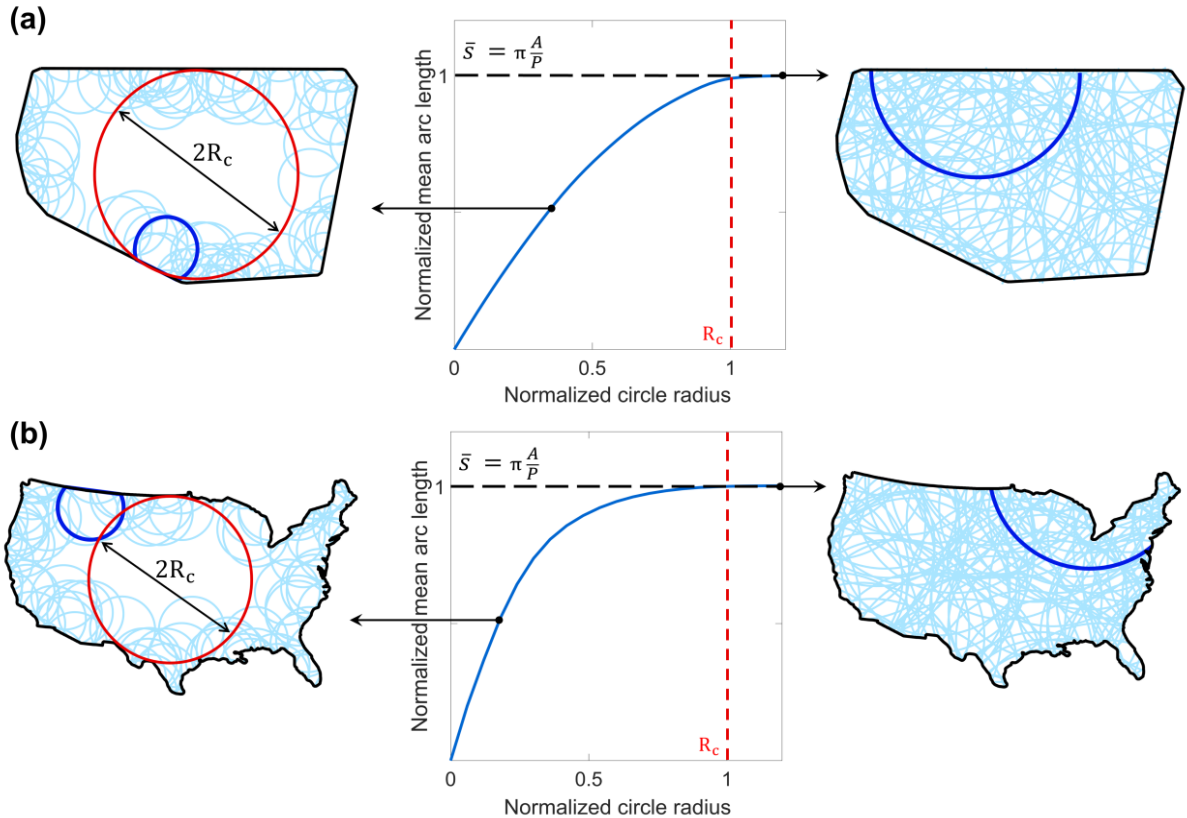

**Figure 1 Mean arc length for circular trajectories intersecting convex and non-convex domains.** Evolution of the normalized mean arc length $\bar{s}$ as a function of the normalized trajectory radius $R/R_c$, $R_c$ being the radius of the largest inscribed circle in the explored domain (red circle). Blue circles exemplify circular trajectories intersecting the domain boundaries (a) the domain is the convex hull of the USA map and (b) the domain is the non-convex map of the USA. Insets: examples of associated realizations for $R < R_c$ (left panel) and $R > R_c$ (right panel). The simulations are performed with $10^5$ circles with a uniform spatial distribution and we only consider those intersecting the boundaries of the domains. The horizontal dashed line represents the generalized Cauchy formula $\bar{s} = \pi A/P$, $A$ and $P$ being the area and perimeter of the explored domain, respectively. The mean arc length is normalized by $\pi A/P$.

We start by studying the special case of circular trajectories. The analysis of circular orbits is of particular interest since circular motion is ubiquitous and represents the simplest closed trajectory. Its rotational invariance avoids integration over all possible orientations of the trajectory which significantly decreases computation time. Moreover, it corresponds to possible existing trajectories such as electrons in a constant magnetic field, swimming microorganisms (20) or robots (23).

We first investigate the case of convex 2D domains. Figure 1a shows the evolution of the mean arc length $\bar{s}(R)$ as a function of the radius $R$ of circular trajectories randomly intersecting the boundary of a 2D convex domain of perimeter $P$ and area $A$ (in this case, the convex hull of the USA map). The trajectories are selected from an initially homogeneous distribution of circles in space and only those intersecting the domain boundary $\partial\Omega$ are retained. Numerical simulations are performed with $10^5$ trajectories of that type. Insets illustrate the domain $\Omega$ (black solid line) and examples of intersecting arc lengths (blue) for trajectories of a fixed radius $R$ smaller than the largest inscribed circle $R_c$ (red circle) and one with trajectories of radius $R > R_c$.

The mean arc length $\bar{s}(R)$ increases steadily from zero in the limit $R \to 0$ reaching a plateau above a critical radius $R_c$ which we observe to correspond to the largest possible inscribed circle in the domain $\Omega$. For $R \geq R_c$, the mean arc length is found to satisfy $\bar{s}(R) = \pi A/P$ (horizontal dashed line). For the asymptotic value of large radius $R \to +\infty$, this value agrees with Cauchy's theorem.

We proceed by extending this investigation to non-convex domains. Figure 1b shows a domain in the shape of the map of the USA. $\bar{s}(R)$ appears to follow the same qualitative evolution as in the case of the convex domain with a continuous increase from zero in the limit $R \to 0$ up to a critical radius $R_c$ where, again, $\bar{s}(R)$ is found to satisfy $\bar{s}(R) = \pi A/P$. The critical radius, as in the convex domain case, is also found to correspond to the largest inscribed circle. The asymptotic solution for infinite $R$ when circular arcs become straight chords agrees with Cauchy's theorem for non-convex domains (25).

These simulations suggest that Cauchy's theorem can be generalized to circular trajectories and even to more complex types of trajectories for arbitrary domains, whether convex or not. They also indicate the existence of a critical size for the trajectory below which the value of the average arc length decreases. This critical size appears to be linked to the possibility of having a trajectory completely included in the domain. We propose now to rationalize these observations.

## 3. Proof of the mean arc length theorem for arbitrary closed trajectories

Objects and properties in integral geometry are quantified by means of the kinematic $dK$-measure (2). It is a measure invariant under rigid transformations. In the following, we designate the explored bounded domain $\Omega_1$ with an area $A_1$ and perimeter $P_1$ and the trajectory by $\partial\Omega_2$ defined as the contour of a second domain $\Omega_2$ of area $A_2$ and perimeter $P_2$. We restrict the topology of $\Omega_1$ to simple connected domains without holes.

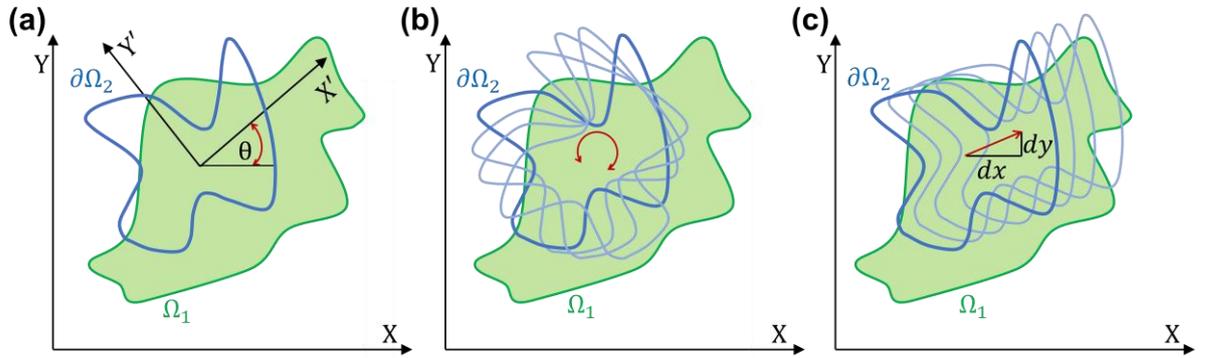

**Figure 2 Schematic representation of the measure of a domain $\Omega_1$ by rigid transformed replicas of a contour $\partial\Omega_2$.** (a) The region $\Omega_1$ (light green) is fixed in the frame XY while the contour $\partial\Omega_2$ (dark blue) is placed in a moving frame X'Y' which can perform a rigid transformation relative to XY. Replicas of $\partial\Omega_2$ (light blue) are obtained (b) by rotations $d\theta$ and (c) by translations along the vector $(dx, dy)$.

A region $\Omega_1$ is fixed in the frame XY while a contour $\partial\Omega_2$ is placed in a moving frame X'Y', that can perform rigid transformations in relation to XY in rotation $d\theta$ or translation along $(dx, dy)$ (see Fig. 2). The kinematic measure of a function $f$ relating the domain $\Omega_1$ and the replicas of the contour $\partial\Omega_2$ is defined by the integral $\int f(\Omega_1, \partial\Omega_2) dK$ over all the rigid transformations of $\partial\Omega_2$, with $dK = dx dy d\theta$ being the kinematic density. The integration is restricted to the situations in which whether $\partial\Omega_2$ or $\Omega_2$ overlaps with $\Omega_1$ which we shall specify further.

Let us recall two central theorems of integral geometry. The first theorem is the *Blaschke's formula* which gives the kinematic measure of the length of congruent contours inside a domain

(26). The kinematic measure of the arc length $s(\Omega_1 \cap \partial\Omega_2)$ that lies inside of the domain $\Omega_1$ for a replica of the contour $\partial\Omega_2$ is given in (27) and is equal to

$$S = \int s(\Omega_1 \cap \partial\Omega_2)dK = 2\pi A_1 P_2 \tag{1}$$

The integration is realized over all points in $\Omega_2$, for all possible replica of $\Omega_2$ such that $\partial\Omega_2 \cap \Omega_1 \neq \emptyset$. The second theorem, called *Poincare's formula*, refers to the number of intersections between an arbitrary curve of finite length and the homogeneous and isotropic replicas of another arbitrary curve (28). The kinematic measure of the intersections $n$ of the contour of domain $\Omega_1$, $\partial\Omega_1$, with the replicas of the contour $\partial\Omega_2$ is given by

$$N_i = \int n(\partial\Omega_1 \cap \partial\Omega_2)dK = 4P_1P_2 \tag{2}$$

The integration is realized over all point in $\Omega_2$, for all possible replica of $\Omega_2$ such that $\partial\Omega_2 \cap \partial\Omega_1 \neq \emptyset$.

In the case of closed contours $\partial\Omega_2$, $n$ is an even number for each intersecting arc. Hence, $N_i$ is twice the number of intersecting arcs.

Provided that no contour is fully included in the domain, which would have no crossing with the boundary of $\partial\Omega_1$, $N_i/2$ is exactly the number of arcs that are involved in eq. 1 to compute $S$. Hence, when this condition is fulfilled, the mean value arc length $\bar{s}$ for closed contour trajectories is thus given by the ratio of $S$ (eq. 1) over $N_i/2$ (eq. 2). It writes

$$\bar{s} = 2\frac{S}{N_i} = \frac{\pi A_1}{P_1} \tag{3}$$

This is a generalized Cauchy formula valid for any arbitrary closed trajectory exploring any arbitrary bounded domain whether convex or not, provided no trajectory can be fully included in the domain. This formula is in agreement with the formula derived by Mazzolo using a similar approach based on integral geometry for random curves (29).

This corresponds precisely to the results presented for circular trajectories exploring domains in Fig. 1. The generalized Cauchy's formula becomes invalid when the radius $R$ of the trajectories are smaller than the maximum inscribed radius $R_c$. It is interesting to note that this demonstration also applies to the case of ballistic straight chords since a bounded domain has always an even number of intersections with a line. This is true whether the domain is convex or not. However, in the case of straight line, there is no size limitation as for the case of closed contours since the number of intersections is never zero when $\partial\Omega_2 \cap \Omega_1 \neq \emptyset$.

# 4. Numerical implementation of the generalized mean arc length theorem

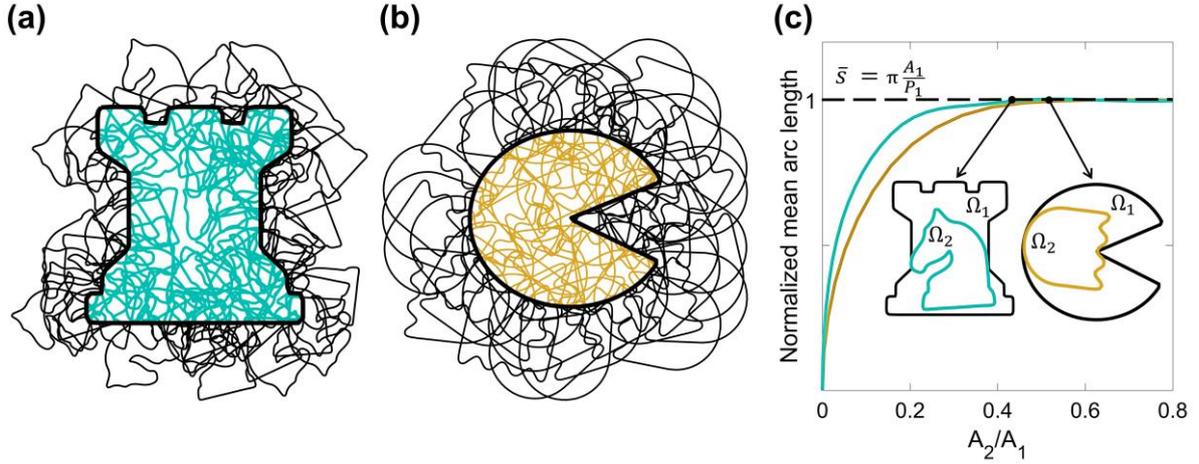

**Figure 3 Mean arc lengths of arbitrary closed trajectories in arbitrary non-convex domains.** Examples of exploration of domains in the shape of (a) the chess Rook and (b) a Pac-Man with trajectories in the shape of a chess Knight and a Ghost respectively. (c) Evolution of the normalized mean arc length as a function of the relative areas of the two domains $A_2/A_1$ for two systems (a) in light blue and (b) in yellow. Inset figures show the largest trajectories completely inscribed in the domain and the associated situation on the curves. The horizontal dashed line represents the generalized Cauchy formula $\bar{s} = \pi A_1/P_1$ with $A_1$ and $P_1$ being respectively the area and perimeter of the explored domain. The simulations are performed with $10^5$ circles randomly intersecting the boundaries of the domains. The mean arc length is normalized by $\pi A_1/P_1$.

We perform numerical simulations to test this generalized Cauchy theorem with complex closed trajectories randomly crossing non-convex regions. Figures 3a and 3b show two domains $\Omega_1$ in the shape of a chess Rook and Pac-Man respectively crossed by randomly distributed trajectories $\partial\Omega_2$ having the shape of a chess Knight and of a Ghost, respectively. Figure 3c shows the evolution of the normalized mean arc length $\bar{s}$ in the two cases as a function of their relative surface $A_2/A_1$. The evolution of $\bar{s}$ with $A_2/A_1$ is qualitatively similar to that obtained with circular trajectories (see Fig. 1) with a steady increase with increasing $A_2/A_1$ before reaching a plateau above a critical value of $A_2/A_1$. The plateaus are reached when all the trajectories intersect the boundaries of the domains, and no trajectory can be fully inscribed inside the domains (see insets in Fig. 3c). This agrees with the limits found in the demonstration and the generalized Cauchy formula (eq. 3).

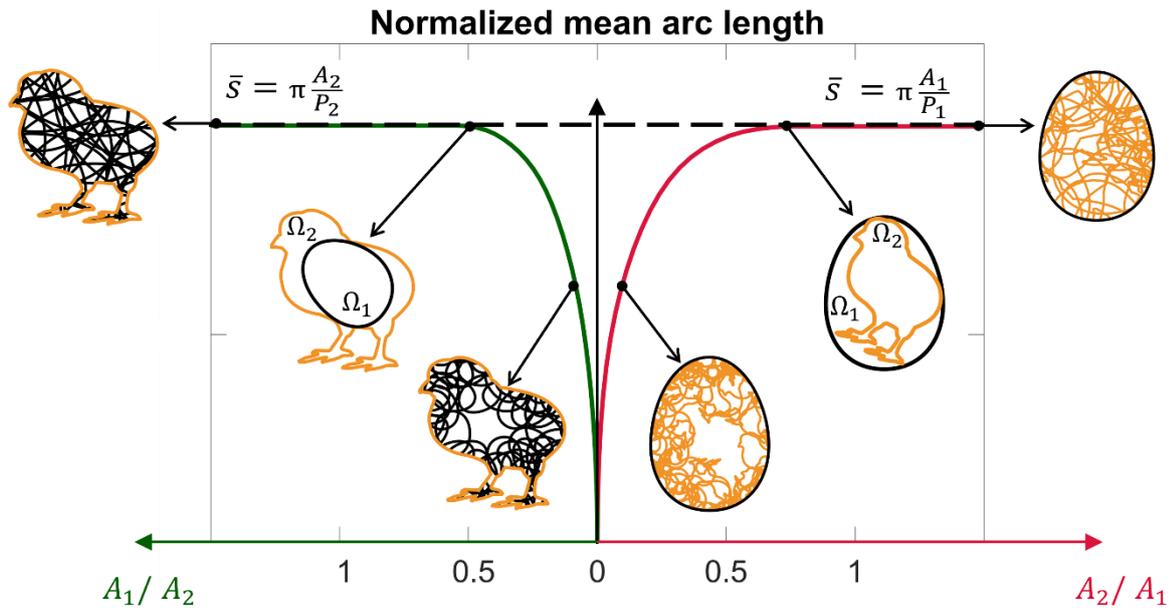

**Figure 4 Reciprocal exploration of the domains**. Mirrored graphs of the evolution of the normalized mean arc length as a function of the relative area for domains in the shape of an egg $\Omega_1$ and a chicken $\Omega_2$: $\Omega_2$ explored by random contours $\partial\Omega_1$ (left) and $\Omega_1$ explored by random contours $\partial\Omega_2$ (right). Insets: examples of realizations in the two asymptotic regimes and the largest contour fully included in the explored domain found by simulations. The horizontal dashed line represents the generalized Cauchy formula $\bar{s}$ normalized by $\pi A_i/P_i$ with $A_i$ and $P_i$ respectively the area and perimeter of the explored domain ($i = 1,2$). The simulations are performed with $10^5$ circles randomly intersecting the boundaries of the domains to be explored.

The role of the random contours and the explored domain can be inter-changed since both boundaries are closed. Figure 4 shows the case of two domains in the shape of an egg (domain $\Omega_1$) and a chicken (domain $\Omega_2$). The mirror graphs plot the evolution of the normalized mean arc length as random egg-shaped trajectories explore a chicken domain as a function of their relative area $A_1/A_2$ (left) and its counterpart, random chicken trajectories explore an egg domain as a function of $A_2/A_1$ (right). Both curves are qualitatively similar to the ones presented previously (Figures 1 and 3) with a steady increase from zero to a plateau satisfying the generalized Cauchy's formula $\bar{s} = \frac{\pi A_i}{P_i}$, $A_i$ and $P_i$ being the area and the perimeter of the explored domain $i = 1,2$, respectively. The plateau values as well as the critical values are a priori different for the two reciprocal cases. The simulations show that these critical values match with the situation of maximum fully inscribed contours in the explored domains (see insets in fig. 5). Based on their relative sizes, this allows the extraction of global geometric information of the egg and chicken domains by statistical measurements with the alternative shape.

## 5. Extending the limits of the mean arc length formula

Interesting analytical developments can even be obtained in the size regime in which the generalized Cauchy's formula is not valid, i.e. when some trajectories are entirely included in the explored domain. These developments are restricted to convex shapes for the trajectories and the explored domain.

Under these assumptions, the Santaló's formula applies (2). It gives the kinematic measure of the number of common areas for two convex domains. For uniformly distributed $\Omega_2$, this is given by the measure of the number of domains $\Omega_1 \cap \Omega_2$

$$N_{\text{tot}} = \int_{\Omega_1 \cap \Omega_2} dK = 2\pi(A_1 + A_2) + P_1 P_2 \tag{4}$$

The integration is realized over all points in $\Omega_2$, for all possible replicas of $\Omega_2$ such that $\Omega_2 \cap \Omega_1 \neq \emptyset$. $N_{\text{tot}}$ is the sum of three types of domains $N_{\text{tot}} = N_\subset + N_\cap + N_\supset$, $N_\subset$, $N_\cap$ and $N_\supset$ being the measures of the number of domains $\Omega_1 \cap \Omega_2$ for $\Omega_2$ fully included in $\Omega_1$, partially overlapping $\Omega_1$ and fully overlapping $\Omega_1$ respectively. For the $N_\cap$ domains, the contour is composed of a succession of a piece of curves of $\partial\Omega_1$ and $\partial\Omega_2$. In the regime in which the relative size of the domain $\Omega_2$ becomes small compared to that of $\Omega_1$, this contour becomes simply formed by two arcs, one of $\partial\Omega_1$ and one of $\partial\Omega_2$ with two intersecting points. As a consequence, in this limit, $N_\cap$ is thus given by $N_\cap = N_i/2 = 2P_1 P_2$, the second equality resulting from the Poincaré's formula (eq. 2). In this limit, $N_\supset = 0$. Thus, the measure of the domains $\Omega_2$ that are fully included in $\Omega_1$ writes $N_\subset = N_{\text{tot}} - N_\cap = 2\pi(A_1 + A_2) - P_1 P_2$.

The measure of the arc length for all intersecting domains ($N_\subset + N_\cap$) is given by eq. (1) and is equal to $2\pi A_1 P_2$. Moreover, the measure of the arc length for the sole $N_\subset$ domains is given by

$$\int_{N_\subset \text{ domains}} s\, dK = P_2 \int_{N_\subset \text{ domains}} dK = P_2 N_\subset$$

where we use that fully enclosed domains $\Omega_2$ have an arc length equal to $P_2$ to get the first equality. So, the measure of the arc length for the $N_\cap$ domains is given by subtracting these last two measures and yields $2\pi A_1 P_2 - P_2 N_\subset = P_2(P_1 P_2 - 2\pi A_2)$. By normalizing this quantity by the measure of the number of $N_\cap$ domains, we obtain the mean arc length in the limit of small $\Omega_2$ domains

$$\bar{s} = \frac{2\pi A_1 P_2 - P_2 N_\subset}{N_\cap} = \frac{\pi A_1}{P_1} - P_2 \frac{N_\subset}{N_\cap} = \frac{P_1 P_2 - 2\pi A_2}{2P_1} \tag{5}$$

These forms are equivalent expressions. The second form indicates that $\bar{s}$ converges to the Cauchy limit $\pi A_1/P_1$ in the regime where there are no more inscribed close-looped trajectories in the domain 1 ($N_c = 0$), which corresponds to Hadwiger's condition (30). The third form is more adapted to the regime of small domains to illustrate the decrease of the mean arc length toward zero, in the limit $(A_2, P_2) \to (0,0)$ while keeping the domain 1 fixed.

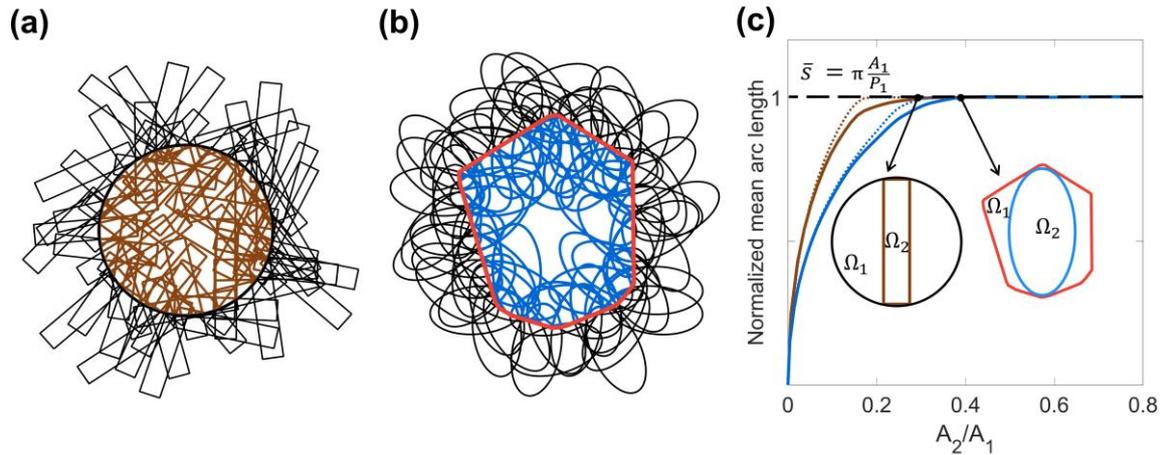

**Figure 5 Extended mean arc length formula for convex domains and trajectories.** Examples of exploration of domains with (a) a disk shape and (b) an arbitrary convex shape with trajectories in the shape of a rectangle and of an ellipse, respectively. (c) Evolution of the normalized mean arc length as a function of the relative areas of the two domains $A_2/A_1$ for systems (a) in red and (b) in blue: simulations with $10^5$ trajectories randomly intersecting the boundaries of the explored domain (solid lines) and approximate theoretical predictions given by eq. (5) and eq. (3) (dotted lines). Insert figures: largest trajectories completely inscribed in the domain with the associated point on the curve. Generalized Cauchy's formula $\bar{s}$ normalized by $\pi A_1/P_1$ (dashed line) with $A_1$ and $P_1$ the area and perimeter of the explored domain, respectively.

We perform numerical simulations with convex domains and trajectories to test the extended theoretical approximate predictions (eq. 5). Figure 5a and 5b show, respectively, a disk-shaped domain explored by randomly distributed rectangular-shaped trajectories and a more complex convex-shaped domain explored with elliptic trajectories. Figure 5c shows the evolution of the normalized simulated arc length (solid line) for different relative sizes $A_2/A_1$ in both cases (red and blue respectively). The analytical formula for $\bar{s}$ is given by eq. (5) for $\bar{s} < \pi A_1/P_1$ and by the generalized Cauchy's formula otherwise. The model (dotted lines) shows a very good agreement with the simulated data with a small discrepancy when the relative size of the trajectories becomes similar to that of the domain. The critical values of $A_2/A_1$ associated to the largest inscribed trajectories in each domain corresponds to the limit between the two validity domains (see inset figures).

Mazzolo (29) recently showed that the ratio between eq. (1) and eq. (4) gives the exact formula for mean arc length obtained from all the contours of $\Omega_2$ included in $\Omega_1$ when the maximal radius of curvature of domain $\Omega_2$ is smaller than the minimal radius of domain $\Omega_1$. Such a formula is related to eq. (5) by removing the contribution of the loops that are entirely contained in $\Omega_1$. Hence, eq. (5) gives a result which is always smaller than Mazzolo's formula. Under the same constraint on the radii of curvature, which insures that the loops $\Omega_2$ cross the boundary of $\Omega_1$ only once, eq. (5) also becomes exact. Interestingly, numerical simulations show that eq. (5) can also be considered an accurate approximation even when this constraint is removed (see Fig. 5c).

## 6. Conclusion

We studied the mean arc length for closed trajectories randomly crossing 2D domains and proved that the mean arc length coincides with the Cauchy's formula for mean chords length under the condition that no trajectory is totally inscribed in the domain. This generalization is valid for arbitrary non-convex trajectories and domains. When this condition is not fulfilled, the mean arc length decreases toward zero as the perimeter of trajectory decreases. In this regime, a simple analytical formula can still describe the evolution of the mean arc length for convex trajectories and domains in the approximation of small trajectories compared to the size of the explored domain. This extension of the mean chord length theorem to arbitrary closed trajectories could find applications in physics, robotics, or biology to retrieve global geometrical information about a domain from the simple measurement of the mean first passage time or length of homogeneously distributed particles entering and exiting this domain.

## Acknowledgements

This project has received funding from the European Union's Horizon 2020 research and innovation program under the Marie Skłodowska-Curie grant agreement No 754387. S. Hidalgo-Caballero thanks CONACYT Mexico for the mobility grant given. The authors thank the support of AXA research fund and the French National Research Agency LABEX WIFI (ANR-10-LABX-24).